\documentclass[]{pasj01}
\draft

\begin{document} 
\Received{}
\Accepted{}

\title{CO($J=3-2$) On-the-fly Mapping of the Nearby Spiral Galaxies NGC~628 and NGC~7793:
Spatially-resolved CO($J=3-2$) Star-formation Law}

 \author{%
   Kazuyuki \textsc{Muraoka}\altaffilmark{1},
   Miho \textsc{Takeda}\altaffilmark{1},
   Kazuki \textsc{Yanagitani}\altaffilmark{1},
   Hiroyuki \textsc{Kaneko}\altaffilmark{2},
   Kouichiro \textsc{Nakanishi}\altaffilmark{3,4},
   Nario \textsc{Kuno}\altaffilmark{5},
   Kazuo \textsc{Sorai}\altaffilmark{6},
   Tomoka \textsc{Tosaki}\altaffilmark{7},
   and 
   Kotaro \textsc{Kohno}\altaffilmark{8}
}

 \altaffiltext{1}{Department of Physical Science, Osaka Prefecture University, Gakuen 1-1, Sakai, Osaka 599-8531}
 \email{kmuraoka@p.s.osakafu-u.ac.jp}
 \altaffiltext{2}{Nobeyama Radio Observatory, Minamimaki, Minamisaku, Nagano 384-1305}
 \altaffiltext{3}{National Astronomical Observatory of Japan, 2-21-1 Osawa, Mitaka, Tokyo 181-8588}
 \altaffiltext{4}{SOKENDAI (The Graduate University for Advanced Studies), 2-21-1 Osawa, Mitaka, Tokyo 181-0015}
 \altaffiltext{5}{Graduate School of Pure and Applied Sciences, University of Tsukuba, 1-1-1 Tennodai, Tsukuba, Ibaraki 305-8577}
 \altaffiltext{6}{Division of Physics, Graduate School of Science, Hokkaido University, Sapporo 060-0810}
 \altaffiltext{7}{Department of Geoscience, Joetsu University of Education, Joetsu, Niigata 943-8512}
 \altaffiltext{8}{Institute of Astronomy, The University of Tokyo, 2-21-1 Osawa, Mitaka, Tokyo 181-0015}


\maketitle

\begin{abstract}
We present the results of CO($J=3-2$) on-the-fly mappings of two nearby non-barred spiral galaxies NGC~628 and NGC~7793
with the Atacama Submillimeter Telescope Experiment at an effective angular resolution of 25$^{\prime \prime}$.
We successfully obtained global distributions of CO($J=3-2$) emission over the entire disks at a sub-kpc resolution for both galaxies.
We examined the spatially-resolved (sub-kpc) relationship between CO($J=3-2$) luminosities ($L^{\prime}_{\rm CO(3-2)}$)
and infrared (IR) luminosities ($L_{\rm IR}$) for NGC~628, NGC~7793, and M~83, and compared it with global luminosities of JCMT Nearby Galaxy Legacy Survey sample.
We found a striking linear $L^{\prime}_{\rm CO(3-2)}-L_{\rm IR}$ correlation over the 4 orders of magnitude,
and the correlation is consistent even with that for ultraluminous infrared galaxies and submillimeter selected galaxies.
In addition, we examined the spatially-resolved relationship between CO($J=3-2$) intensities ($I_{\rm CO(3-2)}$)
and extinction-corrected star formation rates (SFRs) for NGC~628, NGC~7793, and M~83,
and compared it with that for GMCs in M~33 and 14 nearby galaxy centers.
We found a linear $I_{\rm CO(3-2)}-{\rm SFR}$ correlation with $\sim 1$ dex scatter.
We conclude that the CO($J=3-2$) star formation law (i.e., linear $L^{\prime}_{\rm CO(3-2)}-L_{\rm IR}$ and $I_{\rm CO(3-2)}-{\rm SFR}$ correlations)
is universally applicable to various types and spatial scales of galaxies; from spatially-resolved nearby galaxy disks to distant IR-luminous galaxies, within $\sim 1$ dex scatter.

\end{abstract}

\section{Introduction}

Star formation is one of the most fundamental processes in the evolutions of galaxies from nearby objects to high-redshift ones.
The star-formation law, which is a quantitative relationship between star formation rates (SFRs) and surface mass densities of molecular gas,
is represented as a simple power-law and it is often referred to as the Schmidt-Kennicutt law (e.g., \cite{schmidt1959, kennicutt1998a}).
Recently, variations in the power-law index and dispersion of the star-formation law according to the differences in star-formation environments
and physical/chemical properties of molecular gas have been investigated by several authors on the basis of low-$J$ CO observations.
For example, \citet{daddi2010} suggested two different star formation modes in the gas mass versus SFR plane.
One is a rapid starburst mode appropriate for ultraluminous infrared galaxies (ULIRGs), submillimeter selected galaxies, and local starbursts,
and the other is a long-lasting mode appropriate for normal galaxy disks.
The former mode shows about 1 dex higher star formation efficiencies (SFEs, defined as SFR per unit gas mass) than the latter mode.
Such bimodal behavior of the star-formation law may be due to the effects of
a top-heavy initial mass function in starbursts and/or the difference in dense gas fraction (e.g., \cite{miura2014} and references therein).
\citet{krumholz2012} provided a theoretical explanation for such a disk-starburst bimodality.
The authors showed that their sample from Galactic clouds to submillimeter galaxies all lie on a single star formation law
in which the SFR is simply $\sim 1\%$ of the molecular gas mass per local free-fall time.
\citet{leroy2013} demonstrated a first-order linear correspondence between surface densities of molecular gas and SFRs but also found
second-order systematic variations for 30 nearby galaxies at a spatial resolution of 1 kpc.
The authors found that the apparent molecular gas depletion time, which is an inverse of SFE,
decreases with the decrease in stellar mass, metallicity, and dust-to-gas ratio.
This can be explained by a CO-to-H$_2$ conversion factor that depends on dust shielding.

In order to understand the relationship between molecular gas and SFR further, higher-$J$ CO transition gives an important clue
because it can directly trace star-forming denser molecular medium owing to its high critical density.
For instance, the star-formation law based on CO($J=3-2$) emission (its critical density for collisional excitation is $\sim 10^4$ cm$^{-3}$) is often reported.
Single-pointing CO($J=3-2$) emission observations toward central regions of nearby galaxies or entire disks of distant galaxies
showed the nearly linear correlations between CO($J=3-2$) line intensities and SFRs
(or between CO($J=3-2$) luminosities, $L^{\prime}_{\rm CO(3-2)}$, and infrared (IR) luminosities, $L_{\rm IR}$)
with a better correlation coefficient than those based on CO($J=1-0$) emission (e.g., \cite{narayanan2005, komugi2007, iono2009, mao2010}),
which suggests that measurements of CO($J=3-2$) intensities correspond to a simple count of star-forming dense cores within the observing beam
(see also \cite{greve2014} and \cite{liu2015} for similar CO to $L_{\rm IR}$ relations in higher-$J$ transitions).

Obtaining wide-area and spatially-resolved CO($J=3-2$) maps of galaxies is also important
to examine the relationship between dense molecular gas and star formation in galaxy disks.
An extensive CO($J=3-2$) imaging survey of nearby galaxies is conducted using James Clerk Maxwell Telescope (JCMT).
For example, \citet{wilson2009} performed the CO($J=3-2$) mapping of some members of Virgo Clusters
(NGC~4254, NGC~4321, and NGC~4569), and examined a spatial variations in the gas properties
traced by CO($J=3-2$)/CO($J=1-0$) intensity ratio (hereafter $R_{3-2/1-0}$).
The authors found that NGC~4254 has a remarkably uniform $R_{3-2/1-0}$ of 0.33,
whereas NGC 4569 shows a significant gradient from north to south in $R_{3-2/1-0}$;
from 0.53 at the northern CO($J=3-2$) peak to just 0.06 at the southern peak.
\citet{warren2010} presented the CO($J=3-2$) mapping of three nearby field galaxies (NGC~628, NGC~3521, and NGC~3627),
and found that SFE of the dense molecular gas traced by CO($J=3-2$) emission (i.e., $\propto$ SFR divided by $L^{\prime}_{\rm CO(3-2)}$)
mostly independent or only weakly dependent on molecular gas density, $R_{3-2/1-0}$, and the fraction of total gas in molecular form.
\citet{wilson2012} examined the correlation between global $L^{\prime}_{\rm CO(3-2)}$ and far-IR (FIR) luminosities ($L_{\rm FIR}$) for more than 30 galaxies.
The authors found a remarkably tight $L^{\prime}_{\rm CO(3-2)}-L_{\rm FIR}$ correlation among IR-luminous galaxies
(i.e., $L_{\rm FIR}$/$L^{\prime}_{\rm CO(3-2)}$ ratios are constant among IR-luminous galaxies),
whereas $L_{\rm FIR}$/$L^{\prime}_{\rm CO(3-2)}$ ratios (and their scatter) tend to increase among the fainter galaxies.

Another extra-galactic CO($J=3-2$) imaging survey of nearby spiral galaxies has been made
using the Atacama Submillimeter Telescope Experiment (ASTE: \cite{ezawa2004, ezawa2008}),
and obtained sensitive CO($J=3-2$) images of M~83 \citep{muraoka2007, muraoka2009},
M~33 \citep{tosaki2007, miura2012, miura2014}, and NGC~986 \citep{kohno2008}.
We found a linear correlation between CO($J=3-2$) intensities
and extinction-corrected H$\alpha$ luminosities over the whole disk of M~83 \citep{muraoka2009}.
However, in order to reveal the ``universal'' star-formation law in galaxies based on CO($J=3-2$) emission further,
it is indispensable to examine its dependence on various galaxy properties and environments,
such as density and temperature of molecular gas, metallicity, and the difference between nuclear starbursts and star-forming regions
in galaxy disks with an adequate spatial resolution ($\leq$ 1 kpc).
We therefore need to increase the number of spatially-resolved studies of the CO($J=3-2$) star-formation law
covering whole disks of nearby galaxies with various galaxy properties and environments. 

In this paper, we present wide-area CO($J=3-2$) images of nearby non-barred spiral galaxies NGC~628 and NGC~7793
using the ASTE, employing an on-the-fly (OTF) mapping mode.
Basic parameters of each galaxy are summarized in table~1.
The distances to NGC~628 and NGC~7793 are estimated to be 7.3 Mpc and 3.91 Mpc \citep{karachentsev2004};
therefore, the effective beam size of ASTE, $25^{\prime \prime}$, corresponds to 900 pc and 480 pc, respectively.
This enables us to resolve major structures (i.e., the center and spiral arms) at a sub-kpc resolution.
NGC~628 and NGC~7793 are suitable targets to examine a spatially-resolved CO($J=3-2$) star-formation law and to compare it with that in M~83
because the distance to M~83 is estimated to 4.5 Mpc \citep{thim2003}, which is similar to NGC~628 and NGC~7793.
And more importantly, the star-formation environment in M~83 is different from these two galaxies;
M~83 hosts a nuclear starburst, while NGC~628 and NGC~7793 are normal disk galaxies without strong nuclear activities.
Thus, we can investigate the difference in the CO($J=3-2$) star-formation law between the starbursts and star-forming regions in galaxy disks.
These galaxies are rich in multi-wavelength data set to calculate SFRs, such as H$\alpha$ and IR images,
since these galaxies are the targets of the $Spitzer$ Infrared Nearby Galaxies Survey (SINGS: \cite{kennicutt2003})
and/or the Local Volume Legacy (LVL) survey project \citep{kennicutt2008, dale2009}.

The goals of this paper are:
(1) to reveal the global distributions of CO($J=3-2$) emission in NGC~628 and NGC~7793,
(2) to measure CO($J=3-2$) intensities ($I_{\rm CO(3-2)}$) and $L^{\prime}_{\rm CO(3-2)}$, and examine $R_{3-2/1-0}$ in these two galaxies,
(3) to examine the spatially-resolved (sub-kpc) CO($J=3-2$) star-formation law
(i.e., $L^{\prime}_{\rm CO(3-2)}-L_{\rm IR}$ and $I_{\rm CO(3-2)}-{\rm SFR}$ correlations) for NGC~628, NGC~7793, and M~83,
and to compare the obtained CO($J=3-2$) star-formation law with earlier studies,
and (4) to investigate the dependence of the CO($J=3-2$) star-formation law on star-formation environments
(i.e., the difference among ULIRGs, submillimeter selected galaxies, local starbursts, and normal star-forming regions in galaxy disks).

\section{Observations and Data Reduction}

CO($J=3-2$) emission observations of NGC~628 and NGC~7793 were performed
using the ASTE 10-m dish from September to October, 2013, and July to August, 2014, respectively.
The sizes of CO($J=3-2$) maps are $6' \times 6'$ (12.8 kpc $\times$ 12.8 kpc) for NGC~628
and $5' \times 5'$ (5.8 kpc $\times$ 5.8 kpc) for NGC~7793.
The mapped area in each galaxy is indicated in figure~1.

We used a waveguide-type sideband-separating SIS mixer receiver for the single side band (SSB) operation, CATS345 \citep{ezawa2008, inoue2008}.
The typical image rejection ratio of CATS345 was estimated to be $\sim 10$ dB at the frequency in which CO($J=3-2$) emission was observed.
The backend we used was a digital autocorrelator system, MAC \citep{sorai2000}, which comprises four banks of a 512 MHz
wide spectrometer with 1024 spectral channels each.
This arrangement provided a velocity coverage of 440 km s$^{-1}$ with a velocity resolution of 0.43 km s$^{-1}$.
Since the observations were carried out in excellent atmospheric conditions
(the zenith opacity of 220~GHz ranged from 0.03 to 0.10),
the system noise temperature was typically $200-300$ K (in SSB).
We performed the OTF mapping along two different directions
(i.e., scans along the R.A. and Decl. directions),
and these two data sets were co-added by the Basket-weave method \citep{emerson1988}.

We periodically observed CO($J=3-2$) emission of M~17~SW and CW Leo to obtain the main beam efficiency $\eta_{\rm MB}$ in each observing run.
We compared our CO($J=3-2$) spectra with those obtained by CSO observations \citep{wang1994},
and $\eta_{\rm MB}$ was estimated to be $0.54-0.60$ for observing runs in 2013 (NGC~628) and $0.60-0.70$ for those in 2014 (NGC~7793).
The absolute error of the CO($J=3-2$) temperature scale was about $\pm$ 20\%,
mainly due to variations in $\eta_{\rm MB}$ and the image rejection ratio of the CATS345.
Observation parameters are summarized in table~2.

The data reduction was conducted using the software package NOSTAR, which comprises tools for OTF data analysis,
developed by NAOJ \citep{sawada2008}. The raw data were regridded to 7$^{\prime \prime}$.5 per pixel,
giving an effective angular resolution of $\sim 25^{\prime \prime}$.
Linear baselines were subtracted from the spectra. In addition, we subtracted third-order baseline
from a portion of the spectra whose line width are sufficiently narrow ($<$ 40 km s$^{-1}$),
which do not influence line profiles of CO($J=3-2$) emission.
We binned the adjacent channels to a velocity resolution of 5 km s$^{-1}$ for the CO($J=3-2$) spectra.
Finally, 3D data cubes were created.
The resultant r.m.s.\ noise level (1 $\sigma$) in $T_{\rm MB}$ scale at a beam size of 25$^{\prime \prime}$ (HPBW)
was typically $\sim 25$ mK and $\sim 11$ mK for NGC~628 and NGC~7793, respectively.

\section{Results}

\subsection{CO($J=3-2$) Channel Maps, Intensities, and Luminosities}

The derived velocity channel maps of CO($J=3-2$) emission in NGC~628
and NGC~7793 are shown in figure~2 and figure~3, respectively.
No strong concentration of CO($J=3-2$) emission toward central regions are found in either galaxy,
which is different from previous our CO($J=3-2$) sample, barred spiral galaxies M~83 and NGC~986, showing strong peaks at their centers.

We calculated the velocity-integrated intensities of CO($J=3-2$) emission (i.e., $I_{\rm CO(3-2)}$) in NGC~628 and NGC~7793
with a noise cut-off level of 2$\sigma$ and created their maps as shown in figure~4.
We successfully obtained the global CO($J=3-2$) distributions at a sub-kpc resolution for both galaxy disks;
we note that this is the first global CO image of NGC~7793 at all the transitions.
The CO($J=3-2$) image of NGC~628 is similar to that obtained with the JCMT \citep{warren2010, wilson2012},
yet our map depicts even CO($J=3-2$) emission extending southward along the spiral arm.
We compare the distributions of CO($J=3-2$) emission with $Herschel$/SPIRE 350 $\micron$ emission
obtained by the KINGFISH (Key Insights on Nearby Galaxies: a Far-Infrared Survey with $Herschel$) project \citep{kennicutt2011}.
In figure~5, we find a good spatial coincidence between CO($J=3-2$) emission and 350 $\micron$ emission for each galaxy,
which suggests the coexistence of the dense molecular medium and the cold dust component.

We summarize the peak intensity of CO($J=3-2$) emission and its location in each galaxy.
For NGC~628, we found the peak intensity in our CO($J=3-2$) map of 2.7 $\pm$ 0.5 K km s$^{-1}$ in $T_{\rm MB}$ scale at 25$^{\prime \prime}$ resolution.
This peak value occurs at several positions across the map, including the nucleus as well as several local peaks along spiral arms. 
According to \citet{warren2010}, the strongest $I_{\rm CO(3-2)}$ observed with the JCMT is 3.7 $\pm$ 0.5 K km s$^{-1}$ at 15$^{\prime \prime}$ resolution.
These two values of $I_{\rm CO(3-2)}$ seem in good agreement if we consider the difference in the angular resolution
and the extended distribution of CO($J=3-2$) emission.
For NGC~7793, we found the peak intensity in our CO($J=3-2$) map of 1.5 $\pm$ 0.3 K km s$^{-1}$ in $T_{\rm MB}$ scale at 25$^{\prime \prime}$ resolution,
which is observed not at the center but in the disk. The central $I_{\rm CO(3-2)}$ is only 0.7 $\pm$ 0.1 K km s$^{-1}$.

We have calculated the global and central $L^{\prime}_{\rm CO(3-2)}$, and compared them with those reported in earlier studies.
The $L^{\prime}_{\rm CO(3-2)}$ is calculated as follows:
\begin{eqnarray}
L^{\prime}_{\rm CO(3-2)} = I_{\rm CO(3-2)} \, D^2 \, \Omega \,\,\,\,\,\,\,\,  {\rm K}  \,\, {\rm km} \,\, {\rm s}^{-1} \,\, {\rm pc}^2,
\end{eqnarray}
where $D$ is the distance of a galaxy in pc, and $\Omega$ is the covered area in rad$^2$.
For NGC~628, we found the global $L^{\prime}_{\rm CO(3-2)}$ = (7.1 $\pm$ 1.6) $\times$ $10^7$ K km s$^{-1}$ pc$^2$ over the observed $6' \times 6'$ region.
This value is slightly higher than that obtained by \citet{wilson2012}, 5.2 $\times$ $10^7$ K km s$^{-1}$ pc$^2$.
However, their $L^{\prime}_{\rm CO(3-2)}$ was calculated for a smaller area, $5' \times 5'$.
We recalculated our global $L^{\prime}_{\rm CO(3-2)}$ over the same $5' \times 5'$ region,
and found $L^{\prime}_{\rm CO(3-2)}$ = (6.4 $\pm$ 1.4) $\times$ $10^7$ K km s$^{-1}$ pc$^2$, which is in agreement with that obtained by \citet{wilson2012}.
In addition, we obtained $L^{\prime}_{\rm CO(3-2)}$ within the central 1 kpc in NGC~628 of (4.3 $\pm$ 0.9) $\times$ $10^6$ K km s$^{-1}$ pc$^2$,
and found the ratio of the central $L^{\prime}_{\rm CO(3-2)}$ to the global $L^{\prime}_{\rm CO(3-2)}$ is only 0.061.
This is significantly smaller than that obtained in M~83, 0.22 \citep{muraoka2007}.

For NGC~7793, we found the global $L^{\prime}_{\rm CO(3-2)}$ = (7.4 $\pm$ 1.4) $\times$ $10^6$ K km s$^{-1}$ pc$^2$ over the observed $5' \times 5'$ region,
which is an order of magnitude lower than that in NGC~628.
This is partly because the total H$_{2}$ mass in NGC~7793, $2.0 \times 10^8 M_{\odot}$ (calculated from the CO($J=1-0$) intensity obtained by \cite{israel1995}),
is also an order of magnitude lower than that in NGC~628, $1.51 \times 10^9 M_{\odot}$ \citep{garnett2002}.
In addition, we obtained $L^{\prime}_{\rm CO(3-2)}$ = (4.2 $\pm$ 0.8) $\times$ $10^5$ K km s$^{-1}$ pc$^2$ within the central 1 kpc of NGC~7793,
and found the ratio of the central $L^{\prime}_{\rm CO(3-2)}$ to the global $L^{\prime}_{\rm CO(3-2)}$ of 0.055, which is comparable to that in NGC~628.
Therefore, the contribution of the central $L^{\prime}_{\rm CO(3-2)}$ to the global $L^{\prime}_{\rm CO(3-2)}$ is less than 10 \%
for these two spiral galaxies without strong nuclear activity.

We examined the correlation between global $L^{\prime}_{\rm CO(3-2)}$ and $L_{\rm FIR}$ for NGC~628, NGC~7793, M~83 \citep{muraoka2009}, and galaxies in JCMT Nearby Galaxy Legacy Survey (NGLS) sample \citep{wilson2012}.
We calculated $L_{\rm FIR}$ of ASTE sample (NGC~628, NGC~7793, and M~83) from the original data presented by \citet{sanders2003} for adopted distance.
Figure~6 shows the correlation between $L^{\prime}_{\rm CO(3-2)}$ and $L_{\rm FIR}$ and the variation in $L_{\rm FIR}$/$L^{\prime}_{\rm CO(3-2)}$ as a function of $L^{\prime}_{\rm CO(3-2)}$.
Our ASTE sample reinforces the tight correlation between $L^{\prime}_{\rm CO(3-2)}$ and $L_{\rm FIR}$,
and the declining tendency of $L_{\rm FIR}$/$L^{\prime}_{\rm CO(3-2)}$ with the increase in $L^{\prime}_{\rm CO(3-2)}$ demonstrated by \citet{wilson2012}.

\subsection{CO($J=3-2$)/CO($J=1-0$) Intensity Ratios $R_{3-2/1-0}$}

Here, we examine $R_{3-2/1-0}$ in NGC~628 and NGC~7793 to estimate the average physical condition of molecular gas within the observing beam.

We used the CO($J=1-0$) image in NGC~628 obtained with the Berkeley-Illinois-Maryland Association
millimeter interferometer Survey of Nearby Galaxies (BIMA SONG: \cite{helfer2003}) project,
which is convolved to an angular resolution of 25$^{\prime \prime}$ to match our CO($J=3-2$) image.
We can obtain $R_{3-2/1-0}$ only in the central $\sim 3' \times 3'$ region
due to the smaller coverage of the BIMA CO($J=1-0$) image.
According to the pixel-by-pixel comparison of each CO map, the obtained $R_{3-2/1-0}$ are in the range of 0.1 to 1.1.
We found that $\sim 70$\% of pixels show lower $R_{3-2/1-0}$, $0.10-0.60$, regardless of galactocentric distance,
and found that the global $R_{3-2/1-0}$ over the $\sim 3' \times 3'$ region is estimated to be 0.39,
which are well consistent with those reported by \citet{warren2010} and \citet{wilson2012}.
Note that pixels with higher $R_{3-2/1-0}$ ($\geq$ 1.0) are not adjacent each other, i.e., exist locally.
A similar situation has been observed in M~83; the average $R_{3-2/1-0}$ in the disk ($r > 1$ kpc) of M~83 is $0.6-0.7$,
but there exists some locations where $R_{3-2/1-0}$ exceed 1.2 \citep{muraoka2007}.
Such local peaks of $R_{3-2/1-0}$ are presumably due to the uncertainties and the poor signal-to-noise ratio of both CO data.

We compare the obtained $R_{3-2/1-0}$ in NGC~628 with those in various types of galaxies.
Firstly, we summarize $R_{3-2/1-0}$ in normal spiral galaxies.
The mean $R_{3-2/1-0}$ are $0.4-0.5$ for giant molecular clouds (GMCs) in the disk of the Milky Way \citep{sanders1993, oka2007},
$0.3-0.4$ for NGC~4254 and NGC~4321 \citep{wilson2009}, and $\sim 0.4$ for the spiral arm of M~51 \citep{vlahakis2013}.
These values are almost comparable to the global $R_{3-2/1-0}$ of 0.39 in NGC~628.
In addition, the distribution of spatially-resolved $R_{3-2/1-0}$ in M~51 ($0.1-0.7$: \cite{vlahakis2013}) is
quite similar to that in NGC~628 (typically $0.10-0.60$).
However, higher $R_{3-2/1-0}$ have been frequently observed in starburst galaxies.
For example, the central $R_{3-2/1-0}$ in NGC~253 and M~82 are 0.8 and 1.0, respectively \citep{dumke2001}.
Moreover, \citet{muraoka2007} reported that M~83 shows higher $R_{3-2/1-0}$ both in the central region ($\sim 1.0$) and in the disk ($0.6-0.7$).
This suggests that dense gas fraction traced by $R_{3-2/1-0}$ in starburst galaxies is higher than that in NGC~628.
Lastly, we compare $R_{3-2/1-0}$ in NGC~628 and those in early-type galaxies.
According to \citet{mao2010}, single-pointing $R_{3-2/1-0}$ in early-type galaxies are typically in the range of $0.3-0.5$,
which are comparable to that in NGC~628, while some early-type galaxies (NGC~7077 and NGC~7679) show higher $R_{3-2/1-0}$ ($> 1.0$).
The reason of such a great difference in $R_{3-2/1-0}$ among early-type galaxies is unclear,
but this is presumably due to the difference in the physical condition of molecular gas in each galaxy.

For NGC~7793, there is no available CO($J=1-0$) map, but \citet{israel1995} obtained CO($J=1-0$) spectrum in the central region using the SEST 15-m telescope
at an angular resolution of $43''$.
In order to obtain the central $R_{3-2/1-0}$, we convolved our CO($J=3-2$) data (25$''$) to the same angular resolution (43$''$).
The convolved CO($J=3-2$) spectrum in the central region is shown in figure~7.
The significant emission can be seen in the velocity range of 200 to 300 km s$^{-1}$,
which is consistent with the central H\emissiontype{I} emission obtained with the NRAO Very Large Array \citep{walter2008}.
Within this velocity range, we obtained the central $I_{\rm CO(3-2)}$ at an angular resolution of $43''$ of 0.46 $\pm$ 0.07 K km s$^{-1}$ in $T_{\rm MB}$ scale.
We calculated the central CO($J=1-0$) intensity ($I_{\rm CO(1-0)}$) in the same velocity range as our CO($J=3-2$) emission
based on the CO($J=1-0$) spectrum shown in figure~2 of \citet{israel1995}.
We obtained $I_{\rm CO(1-0)}$ of 2.7 $\pm$ 0.4 K km s$^{-1}$, which gives $R_{3-2/1-0}$ of 0.17 $\pm$ 0.06.

The central $R_{3-2/1-0}$ in NGC~7793, 0.17 $\pm$ 0.06, is considerably lower than those reported by earlier studies
based on single-pointing observations of external galaxies.
For example, \citet{mauersberger1999} found that $R_{3-2/1-0}$ in 28 nearby galaxies are in the range of 0.2 to 0.7,
and \citet{mao2010} reported that $R_{3-2/1-0}$ are in the range of 0.2 to 1.9
for their sample including various types of galaxies; normal, Seyfert, starburst galaxies, and luminous infrared galaxies (LIRGs).
In addition, \citet{meier2001} found that $R_{3-2/1-0}$ in dwarf starburst galaxies are in the range of 0.34 to 2.6,
and $R_{3-2/1-0}$ in compact galaxies are in the range of 0.63 to 1.5 \citep{israel2005}.
However, such a low $R_{3-2/1-0}$ ($<$ 0.20) is sometimes observed as the local minimum in galaxy disks.
In fact, we found the smallest $R_{3-2/1-0}$ of 0.10 in the inter-arm of NGC~628 as described above.
In addition, \citet{onodera2012} found GMCs with $R_{3-2/1-0}$ $\sim 0.1-0.2$ in the second nearest spiral galaxy M~33.
These GMCs with low $R_{3-2/1-0}$ are widely distributed over the disk in M~33,
and their masses are small (typically less than $10^5 M_{\odot}$).
This suggests that such less massive GMCs ($< 10^5 M_{\odot}$) are dominant in the central region of NGC~7793
if we assume the filling factor of CO($J=3-2$) emission is comparable to that of CO($J=1-0$) emission.

Some theoretical models are utilized to obtain the physical interpretation of observed $R_{3-2/1-0}$.
In particular, the Large Velocity Gradient (LVG) approximation \citep{scoville1974, goldreich1974}
and the photo-dissociation region (PDR) models (e.g., \cite{hollenbach1997, hollenbach1999, kaufman1999})
are widely applied to derive physical parameters, such as density ($n_{\rm H_2}$) and kinetic temperature ($T_{\rm K}$) of molecular gas.
For example, if we assume a CO fractional abundance per unit velocity gradient $Z$($^{12}$CO)/($dv$/$dr$)
of $1 \times 10^{-4}$ and a moderate $T_{\rm K}$ of 30 K under the LVG approximation with a one-zone assumption,
the observed $R_{3-2/1-0}$ of 0.39 in NGC~628 and 0.17 in NGC~7793 correspond to low $n_{\rm H_2}$ of
$\sim 10^{2.5}$ cm$^{-3}$ and $\sim 10^{2.1}$ cm$^{-3}$, respectively.
However, a different physical condition can reproduce the same $R_{3-2/1-0}$ value under the LVG approximation;
a moderate $n_{\rm H_2}$ of $10^{3.2}$ cm$^{-3}$ and a low $T_{\rm K}$ of 10 K also yield the $R_{3-2/1-0}$ of 0.17.
Therefore, it is difficult to decide whether a low $R_{3-2/1-0}$ of 0.17 observed in NGC~7793 is attributable to
the decrease in $n_{\rm H_2}$ or that in $T_{\rm K}$.

\section{Discussion}

Validity of a use of CO($J=3-2$) emission as a dense gas tracer, and its correlation with SFRs and (F)IR luminosities
have been energetically studied by many authors.
For example, inclination-corrected $I_{\rm CO(3-2)}$ were compared with extinction-corrected SFRs
for 14 nearby galaxy centers by \citet{komugi2007}.
The authors found the strong correlation between these 2 quantities with a linear slope of 1.0.

The detailed investigation of the $L^{\prime}_{\rm CO(3-2)}-L_{\rm FIR}$ correlation is first performed by \citet{yao2003}.
They found a superlinear slope of 1.4 for 60 IR luminous galaxies.
However, recent studies reported that the reanalysis of \citet{yao2003} data yields a linear slope of 1.0 \citep{mao2010, greve2014}.
In addition, \citet{narayanan2005} found the nearly linear correlation between $L^{\prime}_{\rm CO(3-2)}$ and $L_{\rm IR}$ for 17 starburst spiral galaxies, LIRGs, and ULIRGs.
The obtained $L^{\prime}_{\rm CO(3-2)}-L_{\rm IR}$ slope is 0.92.
Subsequent studies also report nearly linear $L^{\prime}_{\rm CO(3-2)}-L_{\rm FIR}$ (and/or $L^{\prime}_{\rm CO(3-2)}-L_{\rm IR}$) correlations.
The reported slopes are 1.08 for LIRGs, submillimeter selected galaxies, quasars, and Lyman-break galaxies \citep{iono2009},
0.99 for nearby ($D < 10$ Mpc) and high-$z$ ($z \geq 1$) sources \citep{bayet2009},
0.87 for 114 targets including normal, Seyfert, starburst galaxies, and luminous infrared galaxies \citep{mao2010},
and $0.99-1.00$ for a sample of 62 local (U)LIRGs and 35 submillimeter selected dusty star-forming galaxies \citep{greve2014}.

Here, we investigate whether such a linear correlation based on CO($J=3-2$) emission is commonly applicable to each local position in nearby spiral galaxies.
Firstly,  we examine the spatially-resolved (sub-kpc) $L^{\prime}_{\rm CO(3-2)}-L_{\rm IR}$ ($8-1000$ $\micron$) correlation for NGC~628, NGC~7793, and M~83.
We use the CO($J=3-2$) image of M~83 obtained by \citet{muraoka2009}.
In addition, we compare the obtained $L^{\prime}_{\rm CO(3-2)}-L_{\rm IR}$ correlation with global luminosities for JCMT NGLS sample \citep{wilson2012}
and the best-fit relation, log $L_{\rm IR}$ = (1.00 $\pm$ 0.05) log $L^{\prime}_{\rm CO(3-2)}$ + (2.2 $\pm$ 0.5), for the local (U)LIRGs and submillimeter selected dusty star-forming galaxies \citep{greve2014}.
We obtain $L_{\rm IR}$ for JCMT NGLS sample from \citet{sanders2003} for adopted distances.
Second, we examine the spatially-resolved relationship between $I_{\rm CO(3-2)}$ and SFRs
based on extinction-corrected H$\alpha$ luminosities for NGC~628, NGC~7793, and M~83.
We compare the obtained $I_{\rm CO(3-2)}-{\rm SFR}$ correlation with that for GMCs in M~33 and 14 nearby galaxy centers.
We use the single-pointing $I_{\rm CO(3-2)}$ and SFRs of GMCs in M~33 at the spatial resolution of $\sim 100$ pc obtained by \citet{onodera2012}
and those of 14 nearby galaxy centers summarized by \citet{komugi2007}.

\subsection{Correlation between Spatially-resolved $L^{\prime}_{\rm CO(3-2)}$ and $L_{\rm IR}$}

Generally, $L_{\rm IR}$ are calculated from flux densities of multiple IR-bands,
but this makes the estimate of spatially-resolved $L_{\rm IR}$ in nearby galaxy disks difficult.
\citet{dale2009} examined the monochromatic-to-bolometric infrared ratios for globally integrated LVL data.
The authors reported that the $Spitzer$/MIPS 70 $\micron$ and 160 $\micron$ emission are tightly coupled to the bolometric IR emission.
Indeed, the mean 70 $\micron$-to-total IR (TIR; $3-1100$ $\micron$) luminosity ratio is 0.46 with a smaller scatter of 0.11 dex.
Thus, we use MIPS 70 $\micron$ images obtained by the LVL survey to estimate the spatially-resolved $L_{\rm IR}$ for NGC~628, NGC~7793, and M~83.

First, we examined the global 70 $\micron$ luminosity ($L_{70}$) to $L_{\rm IR}$ ratio ($L_{70}$/$L_{\rm IR}$) for each galaxy.
The global 70 $\micron$ luminosities are calculated from the integrated MIPS 70 $\micron$ flux densities summarized by \citet{dale2009},
and the $L_{\rm IR}$ are obtained from \citet{sanders2003}.
The resultant global $L_{70}$/$L_{\rm IR}$ are 0.51, 0.61, and 0.53 for NGC~628, NGC~7793, and M~83, respectively.
These values are slightly greater than the mean ratio of 0.46 reported by \citet{dale2009}.
This is because the wavelength range of their TIR luminosity ($L_{\rm TIR}$; $3-1100$ $\micron$) is wider than that of $L_{\rm IR}$ ($8-1000$ $\micron$),
and thus $L_{\rm TIR}$ is definitely greater than $L_{\rm IR}$ (i.e., $L_{70}$/$L_{\rm IR}$ is greater than $L_{70}$/$L_{\rm TIR}$).
Then, we converted the spatially-resolved $L_{70}$ to $L_{\rm IR}$ for each pixel
by using the global $L_{70}$/$L_{\rm IR}$ of each galaxy as a ``scaling'' factor.
Note that we cannot consider the local variations in $L_{70}$/$L_{\rm IR}$ within each galaxy disk,
and thus we estimated the absolute error of $L_{\rm IR}$ in each pixel of $\pm 30$\%
considering the scatter of $L_{70}$/$L_{\rm TIR}$ of 0.11 dex ($\sim 30$\%) \citep{dale2009}.

For the pixel-by-pixel comparison between $L^{\prime}_{\rm CO(3-2)}$ and $L_{\rm IR}$,
we used pixels whose $L^{\prime}_{\rm CO(3-2)}$ exceed 3 $\sigma$ corresponding to $5.6 \times 10^5$ K km s$^{-1}$ pc$^2$, $8.5 \times 10^4$ K km s$^{-1}$ pc$^2$,
and $7.0 \times 10^5$ K km s$^{-1}$ pc$^2$ for NGC~628, NGC~7793, and M~83, respectively.
In this analysis, we divided M~83 data into two regions according to the galactocentric radius; the central region ($r$ $\leq$ 1 kpc) and the disk ($r >$ 1 kpc),
in order to distinguish the nuclear starburst and star-forming regions in the disk.

Figure~8 shows the obtained $L^{\prime}_{\rm CO(3-2)}-L_{\rm IR}$ correlation.
We found a striking linear correlation over the 4 orders of magnitude.
$L_{\rm IR}$ at a given $L^{\prime}_{\rm CO(3-2)}$ seems typically smaller than
the best-fit relation for the sample of \citet{greve2014} by $0.3-0.5$ dex,
but is almost within the error of the best-fit relation.
Note that we found only weak $L^{\prime}_{\rm CO(3-2)}-L_{\rm IR}$ correlations for NGC~628 and NGC~7793 individually,
with Spearman rank correlation coefficients of 0.31 and 0.23, respectively.
This is presumably due to the lack of the dynamic range in $L^{\prime}_{\rm CO(3-2)}$, only $\sim 0.5$ dex for NGC~628 and NGC~7793,
which is comparable to the scatter in $L_{\rm IR}$ at a given $L^{\prime}_{\rm CO(3-2)}$;
therefore, an individual $L^{\prime}_{\rm CO(3-2)}-L_{\rm IR}$ correlation for each galaxy becomes weaker.
In order to obtain a strong $L^{\prime}_{\rm CO(3-2)}-L_{\rm IR}$ correlation for an individual galaxy,
it is necessary to ensure an enough dynamic range in $L^{\prime}_{\rm CO(3-2)}$ at least an order of magnitude such as M~83.
The important thing is that almost all data points lie on a single $L^{\prime}_{\rm CO(3-2)}-L_{\rm IR}$ correlation in figure~8.

In figure~8, we also found a bimodal correlation for each subsample;
we obtained the best-fit slope for the spatially-resolved sample (NGC~628, NGC~7793, and M~83) of 0.84,
and that for the JCMT NGLS sample of 0.74,
in spite of the linear $L^{\prime}_{\rm CO(3-2)}-L_{\rm IR}$ correlation for the combined sample.
This is because the luminosities (both $L^{\prime}_{\rm CO(3-2)}$ and $L_{\rm IR}$) are
proportional to the coverer area (i.e., $D^2 \Omega$ in equation 1) in a linear scale (pc$^2$).
In fact, $D^2 \Omega$ at sub-kpc resolution of $\sim 10^5$ pc$^2$ is 2 or 3 orders of magnitude smaller than
that for a global galaxy disk of $10^7-10^8$ pc$^2$.
Such a difference in $D^2 \Omega$ among each sample automatically produces the linear $L^{\prime}_{\rm CO(3-2)}-L_{\rm IR}$ correlation
for the combined sample over the 3 orders of magnitude even if the slope of $L^{\prime}_{\rm CO(3-2)}-L_{\rm IR}$ correlation for each subsample is sublinear.

The sublinear $L^{\prime}_{\rm CO(3-2)}-L_{\rm IR}$ slope of 0.74 for the JCMT NGLS sample is consistent with
the declining tendency of $L_{\rm FIR}$/$L^{\prime}_{\rm CO(3-2)}$ ratio with the increase in $L^{\prime}_{\rm CO(3-2)}$ reported by \citet{wilson2012} (see also figure~6).
The authors argued that fainter galaxies have lower average CO surface brightnesses, in which case they could systematically underestimate
the CO luminosity due to the low signal-to-noise ratio.

The $L^{\prime}_{\rm CO(3-2)}-L_{\rm IR}$ slope of 0.84 for our spatially-resolved sample seems
slightly smaller than that reported by earlier studies ($\sim 1.0$).
However, we cannot discuss its significance yet because the slope of 0.84 is determined for only 3 galaxy sample in this study.
Further analyses of the spatially-resolved $L^{\prime}_{\rm CO(3-2)}-L_{\rm IR}$ correlation for more sample are required
to determine more reliable $L^{\prime}_{\rm CO(3-2)}-L_{\rm IR}$ slope and to understand its physical meaning in nearby galaxies.

\subsection{Correlation between Spatially-resolved $I_{\rm CO(3-2)}$ and SFRs}

Extinction-corrected SFRs are calculated from a linear combination of H$\alpha$ and $Spitzer$/MIPS 24 $\micron$ luminosities
as follows \citep{kennicutt1998a, kennicutt1998b, calzetti2007}:
\begin{eqnarray}
{\rm SFR} = 7.9 \times 10^{-42} \left( \frac{L_{{\rm H} \alpha } + 0.031 \times L_{24 \mu {\rm m}}}{{\rm erg} \,\, {\rm s}^{-1}} \right) \frac{{\rm cos} \ i}{\Omega_A} \,\,\,\,\,\,\,\, M_{\odot} \,\, {\rm yr}^{-1} \,\, {\rm pc}^{-2},
\end{eqnarray}
where $L_{{\rm H} \alpha}$ and $L_{24 \mu {\rm m}}$ mean H$\alpha$ and 24 $\micron$ luminosities, respectively.
$i$ is the inclination of each galaxy and $\Omega_A$ is the covered area within the 25$''$ beam (in the unit of pc$^{2}$).
We used H$\alpha$ and 24 $\micron$ images of NGC~7793 obtained by the SINGS and those of M~83 and NGC~628 by the LVL survey.

Figure~9 shows the obtained $I_{\rm CO(3-2)}-{\rm SFR}$ correlation.
We found a linear correlation between $I_{\rm CO(3-2)}$ and extinction-corrected SFRs with $\sim 1$ dex scatter,
whereas we also found a larger scatter ($\sim 2$ orders of magnitude) for GMCs in M~33.
This is presumably due to the small spatial resolution for M~33.
\citet{miura2012} reported that GMCs in M~33 are classified into four types according to their evolutional stages;
i.e., the age of the associated young stellar groups and H\emissiontype{II} regions.
This causes the difference in the estimated SFRs by $2-3$ orders of magnitude for every GMCs.
In addition, the peak positions of CO($J=3-2$) emission for some GMCs are $\sim 100$ pc away from their associated H\emissiontype{II} regions.
This spatial offset just corresponds to the observing beam of CO($J=3-2$) emission.
For these reasons, the larger scatter for GMCs in M~33 in figure~9 is observed.
Nonetheless, the overall $I_{\rm CO(3-2)}-{\rm SFR}$ correlation is still robust even if GMCs in M~33 are included.

Finally, we note the dependence of the obtained CO($J=3-2$) star-formation law on star-formation environments.
As shown in figure~8 and figure~9, the $L^{\prime}_{\rm CO(3-2)}-L_{\rm IR}$ and the $I_{\rm CO(3-2)}-{\rm SFR}$ correlations for the central region of M~83 is
well consistent with those for NGC~628, NGC~7793, and the disk of M~83, and is consistent even with those for other sample in earlier studies
(i.e., ULIRGs, submillimeter selected galaxies, whole disks of JCMT NGLS sample, 14 nearby galaxy centers, and GMCs in M~33).
We conclude that the CO($J=3-2$) star formation law (linear $L^{\prime}_{\rm CO(3-2)}-L_{\rm IR}$ and $I_{\rm CO(3-2)}-{\rm SFR}$ correlations)
is universally applicable to various types and spatial scales of galaxies; from spatially-resolved nearby galaxy disks to distant IR-luminous galaxies, within $\sim 1$ dex scatter.

\section{Summary}

We have performed CO($J=3-2$) emission observations of the $6' \times 6'$ (or 12.8 $\times$ 12.8 kpc at the distance of 7.3 Mpc) region
of the nearby spiral galaxy NGC~628 (M~74) and the $5' \times 5'$ (or 5.8 $\times$ 5.8 kpc at the distance of 3.91 Mpc) region
of the nearby spiral galaxy NGC~7793 with the ASTE at an effective angular resolution of 25$^{\prime \prime}$.
A summary of this work is as follows.

\begin{enumerate}
\item
We successfully obtained global distributions of CO($J=3-2$) emission over the entire disks at a sub-kpc resolution for both galaxies.
In addition, we found that the CO($J=3-2$) emission is well spatially coincident with the $Herschel$/SPIRE 350 $\micron$ emission.

\item
We found that the global $L^{\prime}_{\rm CO(3-2)}$ are (7.1 $\pm$ 1.6) $\times$ $10^7$ K km s$^{-1}$ pc$^2$ for NGC~628
and (7.4 $\pm$ 1.4) $\times$ $10^6$ K km s$^{-1}$ pc$^2$ for NGC~7793, respectively.
We found that no central concentration of CO($J=3-2$) emission;
the ratios of the central ($< 1$ kpc) $L^{\prime}_{\rm CO(3-2)}$ to the global $L^{\prime}_{\rm CO(3-2)}$ are only $\sim 0.06$ for both galaxies.

\item
We found the average $R_{3-2/1-0}$ in NGC~628 of 0.39, which may be a typical value in galaxy disks
according to the comparison with earlier studies for galactic and extragalactic objects.
On the other hand, we found the central $R_{3-2/1-0}$ at 43$''$ resolution in NGC~7793 of 0.17.
Such a low $R_{3-2/1-0}$ was observed for less massive ($< 10^5 M_{\odot}$) GMCs in the disk of M~33,
which suggests that such less massive GMCs are dominant in the central region of NGC~7793.

\item
We examined the spatially-resolved (sub-kpc) $L^{\prime}_{\rm CO(3-2)}-L_{\rm IR}$ correlation for NGC~628, NGC~7793, and M~83,
and compared it with global luminosities of JCMT NGLS sample.
We found a striking linear $L^{\prime}_{\rm CO(3-2)}-L_{\rm IR}$ correlation over the 4 orders of magnitude,
and the correlation is consistent even with that for (U)LIRGs and submillimeter selected galaxies.

\item
We examined the spatially-resolved relationship between $I_{\rm CO(3-2)}$ and extinction-corrected SFRs
for NGC~628, NGC~7793, and M~83, and compared it with that for GMCs in M~33 and 14 nearby galaxy centers.
We found a linear $I_{\rm CO(3-2)}-{\rm SFR}$ correlation with $\sim 1$ dex scatter.

\item
We conclude that the CO($J=3-2$) star formation law (i.e., linear $L^{\prime}_{\rm CO(3-2)}-L_{\rm IR}$ and $I_{\rm CO(3-2)}-{\rm SFR}$ correlations)
is universally applicable to various types and spatial scales of galaxies;
from spatially-resolved nearby galaxy disks to distant IR-luminous galaxies, within $\sim 1$ dex scatter.

\end{enumerate}

\vspace{0.5cm}
We thank the referee for invaluable comments, which significantly improved the manuscript.
The ASTE telescope is operated by National Astronomical Observatory of Japan (NAOJ).
We would like to acknowledge all of the members involved with
the ASTE team for their great efforts in the ASTE project.
This study was financially supported by MEXT Grant-in-Aid for Young Scientists (B) No.\ 24740126.
This work is based on observations made with the Spitzer Space Telescope,
which is operated by the Jet Propulsion Laboratory, California Institute of Technology under a contract with NASA.


\begin{figure}
  \begin{center}
    \includegraphics[width=17cm]{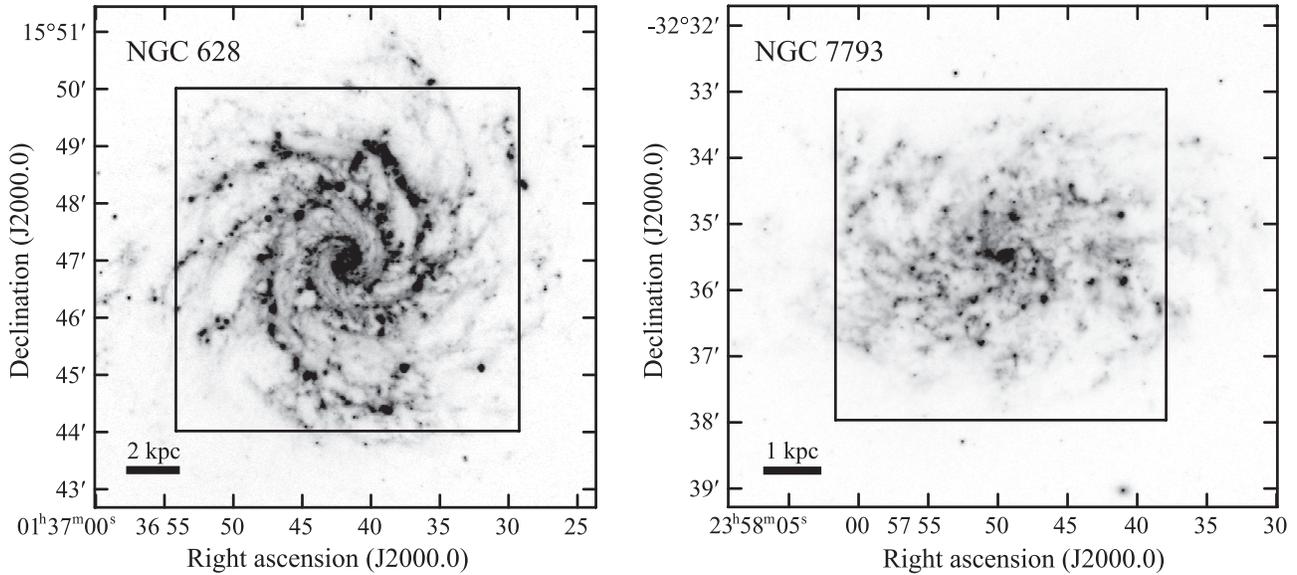}
  \end{center}
\caption{
Observed $6' \times 6'$ area of NGC~628 (left) and $5' \times 5'$ area of NGC~7793 (right), which are indicated by large squares,
superposed on $Spitzer$/IRAC 8 $\micron$ images \citep{dale2009} in order to display footprints of the ASTE observations.
}
\label{fig:fig1}
\end{figure}

\begin{figure}
  \begin{center}
    \includegraphics[width=17cm]{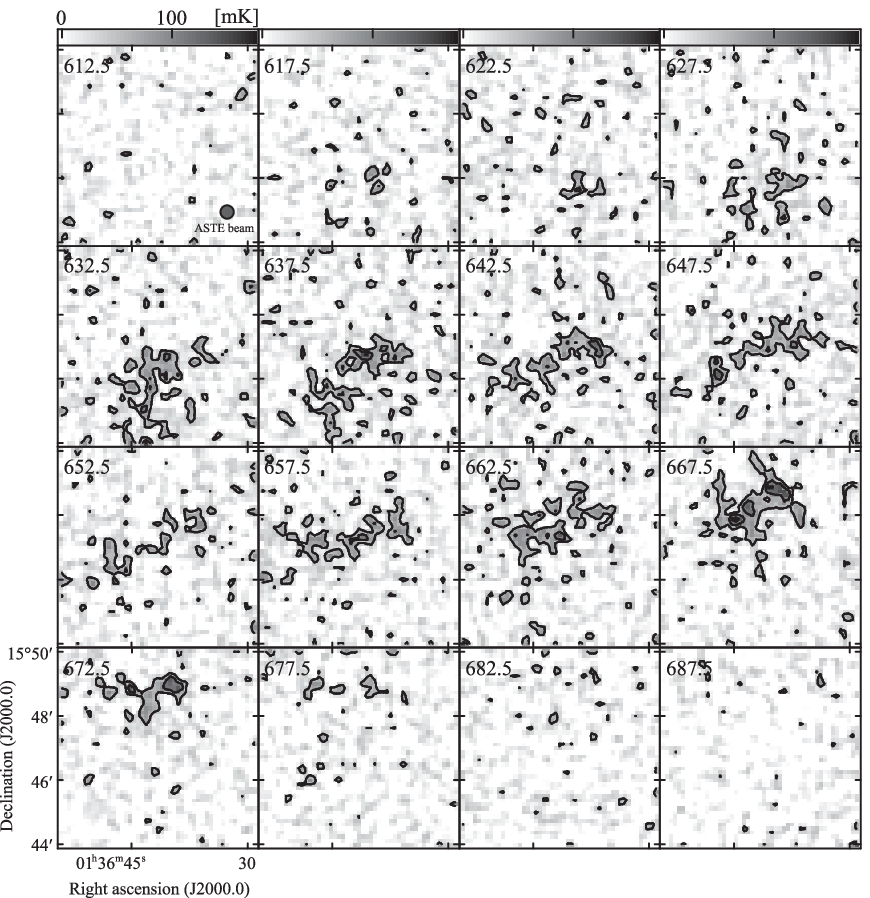}
  \end{center}
\caption{
Velocity channel maps of CO($J=3-2$) emission in NGC~628. The contour levels are 2, 4, and 6$\sigma$,
where 1$\sigma$ = 25 mK in $T_{\rm MB}$ scale. The velocity width of each channel is 5 km s$^{-1}$, and the central velocities ($V_{\rm LSR}$ in km s$^{-1}$)
are labeled in the top-left corner of each map. The beam size of the ASTE (25$''$) is indicated in the bottom-right corner of the first panel.
}
\label{fig:fig2}
\end{figure}

\begin{figure}
  \begin{center}
    \includegraphics[width=17cm]{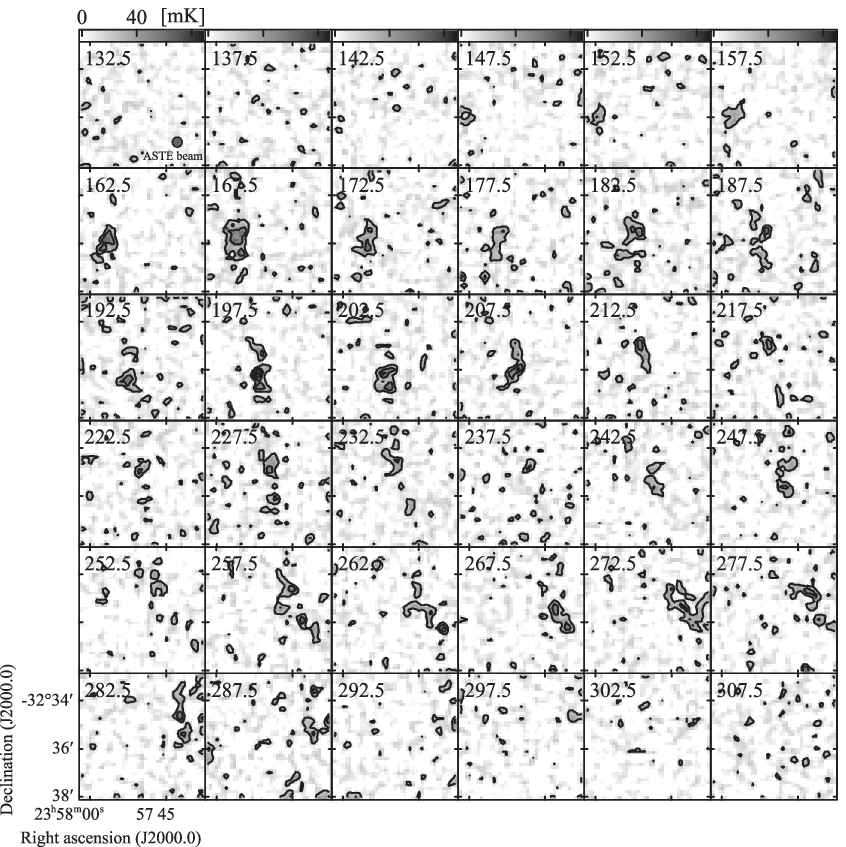}
  \end{center}
\caption{
Velocity channel maps of CO($J=3-2$) emission in NGC~7793. The contour levels are 2, 4, 6, and 8$\sigma$,
where 1$\sigma$ = 11 mK in $T_{\rm MB}$ scale. The velocity width of each channel is 5 km s$^{-1}$, and the central velocities ($V_{\rm LSR}$ in km s$^{-1}$)
are labeled in the top-left corner of each map. The beam size of the ASTE (25$''$) is indicated in the bottom-right corner of the first panel.
}
\label{fig:fig3}
\end{figure}

\begin{figure}
  \begin{center}
    \includegraphics[width=17cm]{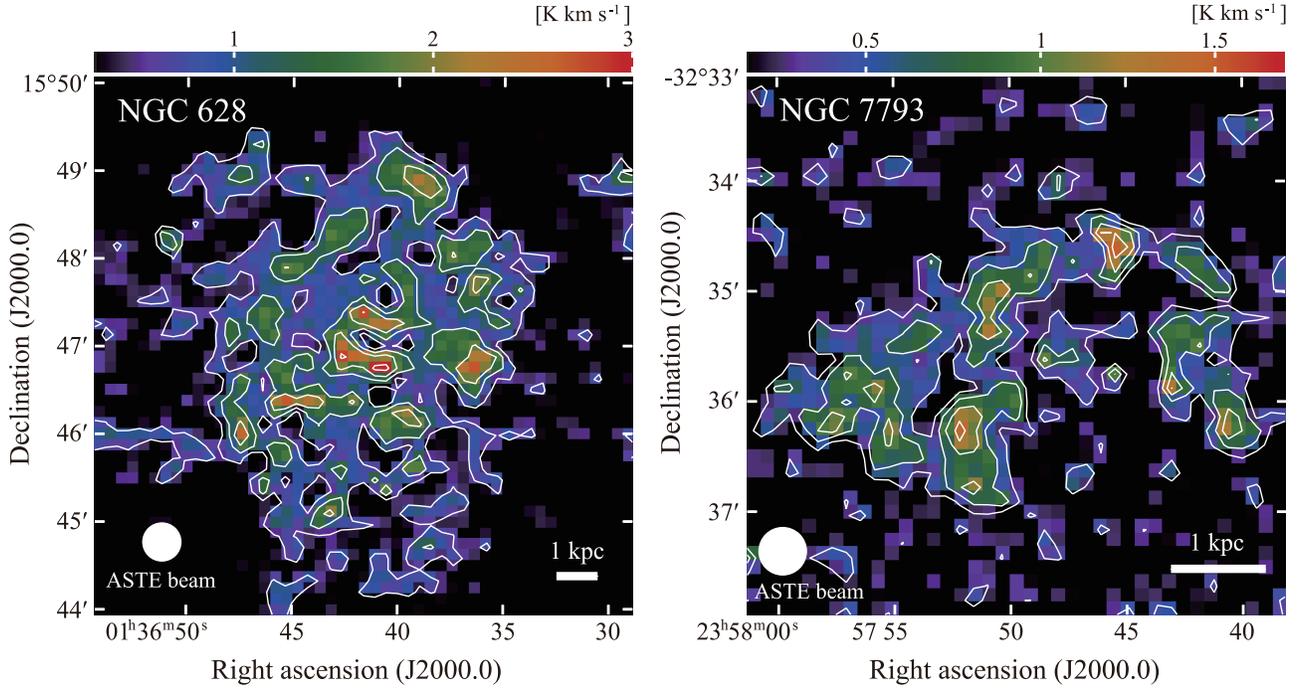}
  \end{center}
\caption{
Maps of velocity-integrated CO($J=3-2$) intensities in NGC~628 (left) and NGC~7793 (right).
The contour levels are 2, 4, 6, and 9$\sigma$, where 1$\sigma$ = 0.30 K km s$^{-1}$ in $T_{\rm MB}$ scale for NGC~628,
and those are 2, 4, 6, and 8$\sigma$, where 1$\sigma$ = 0.16 K km s$^{-1}$ in $T_{\rm MB}$ scale for NGC~7793, respectively.
The beam size of the ASTE (25$''$) is indicated in the bottom-left corner of each map.
}
\label{fig:fig4}
\end{figure}

\begin{figure}
  \begin{center}
    \includegraphics[width=17cm]{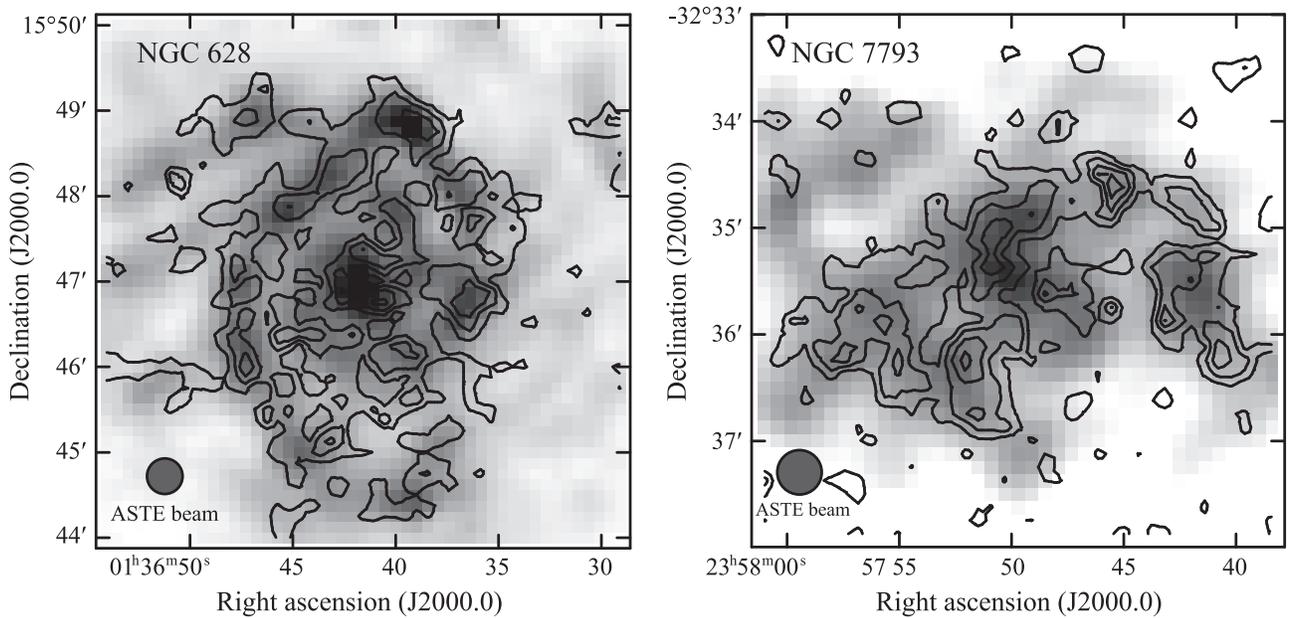}
  \end{center}
\caption{
Maps of velocity-integrated CO($J=3-2$) intensities (contours) superposed on $Herschel$/SPIRE 350 $\micron$ images (gray scale) obtained by \citep{kennicutt2011}
for NGC~628 (left) and NGC~7793 (right). The contour levels are the same as figure 4.
The beam size of the ASTE (25$''$) is indicated in the bottom-left corner of each map.
}
\label{fig:fig5}
\end{figure}

\begin{figure}
  \begin{center}
    \includegraphics[width=17cm]{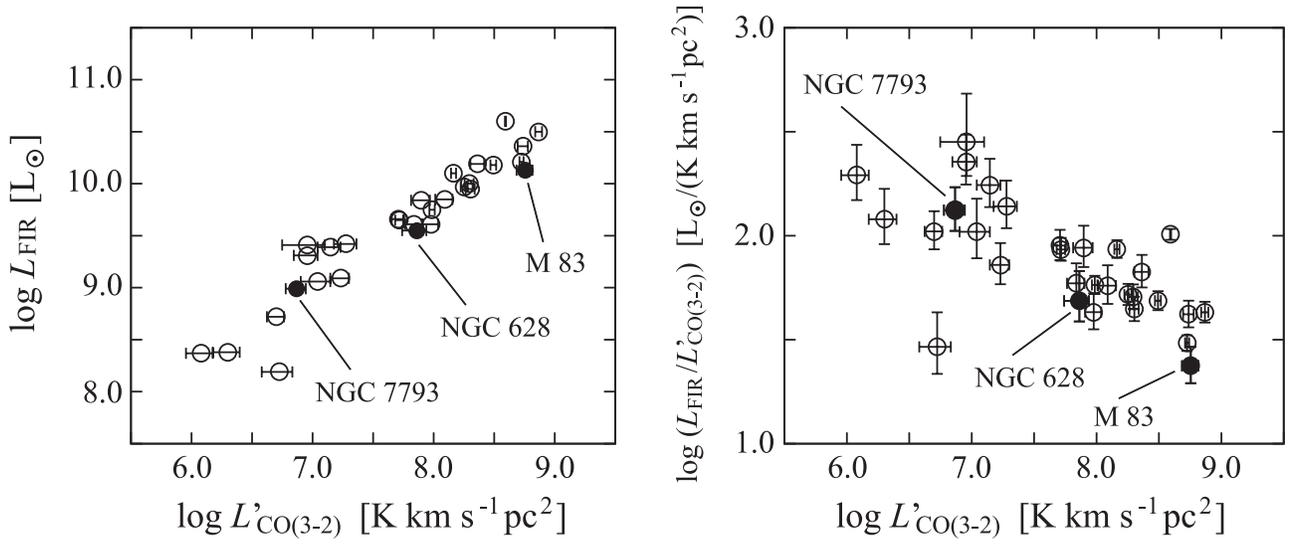}
  \end{center}
\caption{
A correlation between the global $L^{\prime}_{\rm CO(3-2)}$ and $L_{\rm FIR}$ (left)
and the variation in $L_{\rm FIR}$/$L^{\prime}_{\rm CO(3-2)}$ as a function of $L^{\prime}_{\rm CO(3-2)}$ (right)
for our ASTE sample (filled circles) and the JCMT NGLS sample (\cite{wilson2012}; open circles).
}
\label{fig:fig6}
\end{figure}

\begin{figure}
  \begin{center}
    \includegraphics[width=8cm]{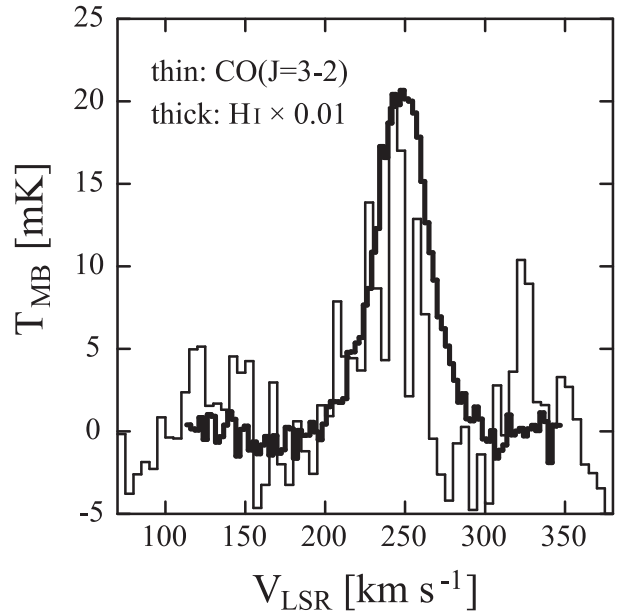}
  \end{center}
\caption{
Spectra of the convolved CO($J=3-2$) emission (thin line) and the H\emissiontype{I} emission (\cite{walter2008}; thick line) at the angular resolution of 43$^{''}$ in the central region of NGC~7793.
The temperature scale of H\emissiontype{I} emission is multiplied by 0.01 for display.
}
\label{fig:fig7}
\end{figure}

\begin{figure}
  \begin{center}
    \includegraphics[width=17cm]{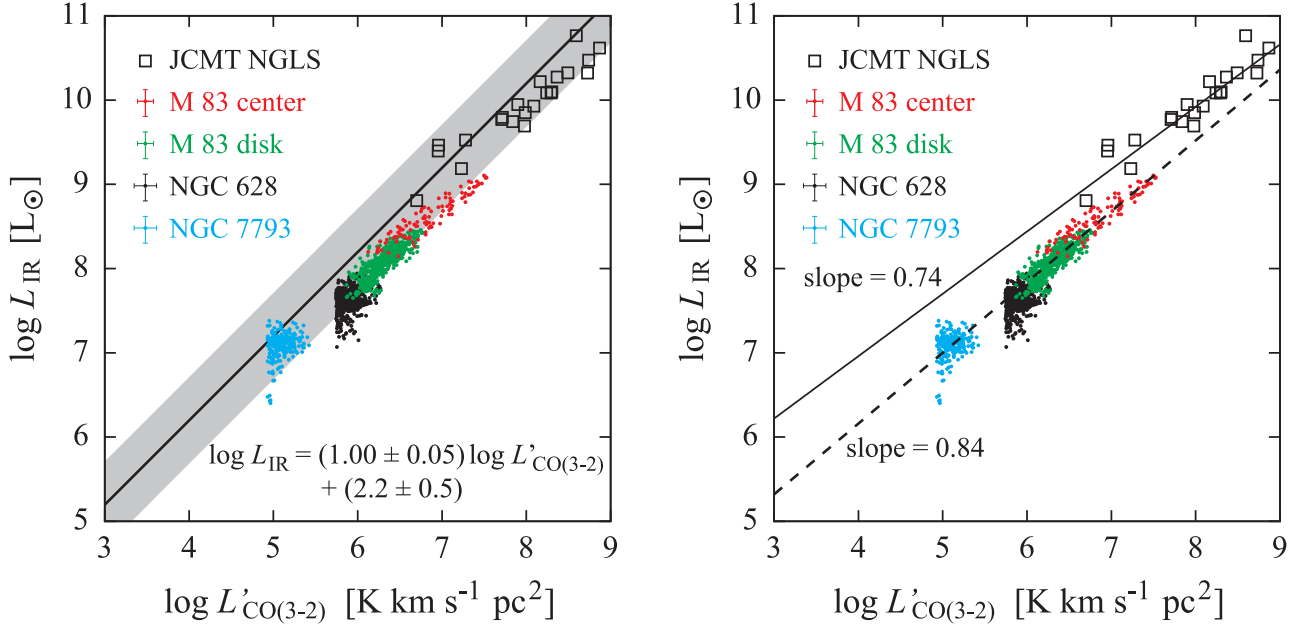}
  \end{center}
\caption{(left) A correlation between $L^{\prime}_{\rm CO(3-2)}$ and $L_{\rm IR}$.
Typical errors due to the calibration uncertainty are shown in the next of each galaxy name.
The solid line and the shaded area indicate the best-fit relation with its error
for the local (U)LIRGs and submillimeter selected dusty star-forming galaxies \citep{greve2014}.
(right) Same as left panel, but the solid line indicates the best-fit slope for JCMT NGLS sample,
and the dashed line indicates that for our spatially-resolved sample.
}
\label{fig:fig8}
\end{figure}

\begin{figure}
  \begin{center}
    \includegraphics[width=8cm]{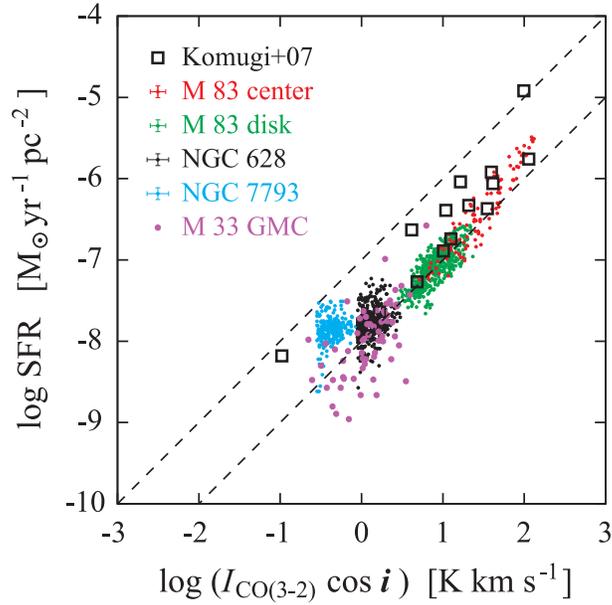}
  \end{center}
\caption{
A correlation between inclination-corrected $I_{\rm CO(3-2)}$ and SFRs.
Typical errors due to the calibration uncertainty are shown in the next of each galaxy name.
Data points of M~33 are derived from CO($J=1-0$) peak positions of individual GMCs \citet{onodera2012}.
Each dashed line indicates a linear correlation (not best-fit for data points) in this plot.
}
\label{fig:fig9}
\end{figure}


\begin{table}
\begin{center}
Table~1.\hspace{4pt}Galaxy parameters\\[1mm]
\begin{tabular}{lll}
\hline \hline \\[-6mm]
Galaxy name & NGC~628 & NGC~7793 \\
\hline \\[-6mm]
Morphological type$^{\rm a}$ & SA(s)c & SA(s)d \\
Map center$^{\rm b}$: &  &   \\[-1mm]
\,\,\,\, Right Ascension  & $1^{\rm h} 36^{\rm m} 41^{\rm s}.7$ & $22^{\rm h} 57^{\rm m} 49^{\rm s}.8$ \\[-1mm]
\,\,\,\, Declination  & $15^{\circ} 47^{\prime} 00^{\prime \prime}.5$ & $-32^{\circ} 35^{\prime} 27^{\prime \prime}.7$ \\
Distance$^{\rm c}$ & 7.3 Mpc & 3.91 Mpc \\
Linear scale & 36 pc arcsec$^{-1}$ & 19 pc arcsec$^{-1}$ \\
Inclination$^{\rm d}$ & 6$^{\circ}$.5 & 53$^{\circ}$.7 \\
12 + log(O/H)$^{\rm e}$ & 8.78 & 8.50 \\
H$_{2}$ mass$^{{\rm f, g}}$ & $1.51 \times 10^9$ $M_{\odot}$ & $2.0 \times 10^8$ $M_{\odot}$ \\
H\emissiontype{I} mass$^{\rm f}$ & $7.43 \times 10^9$ $M_{\odot}$ & $1.05 \times 10^9$ $M_{\odot}$ \\
stellar mass$^{\rm h}$ & $1.48 \times 10^{10}$ $M_{\odot}$ & $3.72 \times 10^9$ $M_{\odot}$ \\
global SFR$^{\rm i}$ & 0.59 $M_{\odot}$ yr$^{-1}$ & 0.30 $M_{\odot}$ yr$^{-1}$\\
specific SFR & $3.99 \times 10^{-11}$ ${\rm yr}^{-1}$ & $8.06 \times 10^{-11}$ ${\rm yr}^{-1}$ \\
$L_{\rm FIR}$ ($4-400$ $\micron$)$^{\rm j}$ & $3.53 \times 10^9$ $L_{\odot}$ & $9.81 \times 10^8$ $L_{\odot}$ \\
$L_{\rm IR}$ ($8-1000$ $\micron$)$^{\rm j}$ & $4.76 \times 10^9$ $L_{\odot}$ & $1.10 \times 10^9$ $L_{\odot}$ \\
\hline \\[-2mm]
\end{tabular}\\
{\footnotesize
$^{\rm a}$Morphological type from RC3.
$^{\rm b}$Map center from \citet{jarrett2003}.
$^{\rm c}$Adopted distances from \citet{karachentsev2004}.
$^{\rm d}$Inclination for NGC~628 from \citet{kamphuis1992} and NGC~7793 from \citet{carignan1985}.
$^{\rm e}$Metallicity form \citet{garnett2002}.
$^{\rm f}$H$_{2}$ mass for NGC~628 and H\emissiontype{I} masses for NGC~628 and NGC~7793 are calculated from the original data presented by \citet{garnett2002} for adopted distances.
$^{\rm g}$H$_{2}$ mass for NGC~7793 is calculated from the central CO($J=1-0$) intensity obtained by \citet{israel1995} assuming the ratio of central to total H$_{2}$ mass being 0.05.
$^{\rm h}$stellar masses are calculated from the equation (8) in \citet{querejeta2015} using $Spitzer$/IRAC 3.6 $\micron$ and 4.5 $\micron$ fluxes presented by \citet{dale2009}.
$^{\rm i}$global SFR from \citet{kennicutt2008}.
$^{\rm j}$$L_{\rm FIR}$ and $L_{\rm IR}$ are calculated from the original data presented by \citet{sanders2003} for adopted distances.
}
\end{center}
\end{table}


\begin{table}
\begin{center}
Table~2.\hspace{4pt}Observation parameters\\[1mm]
\begin{tabular}{lcc}
\hline \hline \\[-6mm]
Galaxy name & NGC~628 & NGC~7793 \\
\hline \\[-6mm]
Observation date & September to October, 2013 & July to August, 2014 \\
Total observation time & 30 hours & 50 hours \\
Field coverage & $6^{\prime} \times 6^{\prime}$ (12.8 $\times$ 12.8 kpc) & $5^{\prime} \times 5^{\prime}$ (5.8 $\times$ 5.8 kpc) \\
System noise temperature & $250-300$ K & $200-250$ K \\
Main-beam efficiency & 0.57 $\pm$ 0.06 $\pm$ 0.03 & 0.65 $\pm$ 0.07 $\pm$ 0.05 \\
Velocity resolution & 5 km s$^{-1}$ & 5 km s$^{-1}$ \\
r.m.s noise level in $T_{\rm MB}$ scale & 25 mK & 11 mK \\
\hline \\[-2mm]
\end{tabular}\\
{\footnotesize
For the main-beam efficiency, the first error indicates systematic error and the second, random error.
}
\end{center}
\end{table}


\end{document}